\documentclass[onecolumn,11pt]{article}
\usepackage[top=1in, bottom=1in, left=1in, right=1in]{geometry}
\usepackage{amsmath,amsfonts,amscd,amssymb}
\usepackage{graphicx}
\usepackage{epstopdf}
\usepackage{overpic}
\usepackage{cancel}
\usepackage{rotating}
\usepackage{url}
\usepackage{caption}
\usepackage{color}
\usepackage{rotating}
\usepackage{multirow}
\usepackage{wrapfig}
\usepackage{mathtools}
\usepackage{subeqnarray}
\usepackage{setspace}
\usepackage{palatino} % needs 10
\usepackage[numbers,sort&compress]{natbib}
\usepackage[bottom,flushmargin,hang,multiple]{footmisc}
\usepackage{lipsum}
\newcommand\blfootnote[1]{%
  \begingroup
  \renewcommand\thefootnote{}\footnote{#1}%
  \addtocounter{footnote}{-1}%
  \endgroup
}

\definecolor{header1}{cmyk}{0,0,0,1}

\title{\vspace{-.25in}{\fontsize{16}{16}\selectfont \textbf{Deep learning to accelerate Maxwell's equations\\ for inverse design of dielectric metasurfaces}}\vspace{-.05in}}
\author{\normalsize{Maksym V. Zhelyeznyakov$^{1}$, Steven L. Brunton$^2$, Arka Majumdar$^{1,3*}$}\\
\footnotesize{$^1$ Department of Electrical and Computer Engineering, University of Washington, Seattle, Washington, 98195, USA}\\
\footnotesize{$^2$  Department of Mechanical Engineering, University of Washington, Seattle, WA 98195, USA}\\
\footnotesize{$^3$ Department of Physics, University of Washington, Seattle, Washington 98195, USA}
}
\date{}
\begin{document}
\maketitle

\blfootnote{$^*$ Corresponding author (arka@uw.edu).}
%%%%%%%%%%%%
%%% ABSTRACT
%%%%%%%%%%%%
\vspace{-.2in}
\begin{abstract}
The inverse design of optical metasurfaces is a rapidly emerging field that has already shown great promise in miniaturizing conventional optics as well as developing completely new optical functionalities. 
Such a design process relies on many forward simulations of a device's optical response in order to optimize its performance. 
We present a data-driven forward simulation framework for the inverse design of metasurfaces that is more accurate than methods based on the local phase approximation, a factor of $10^4$ times faster and requires $15$ times less memory than mesh based solvers, and is not constrained to spheroidal scatterer geometries. 
We explore the scattered electromagnetic field distribution from wavelength scale cylindrical pillars, obtaining low-dimensional representations of our data via the singular value decomposition. 
We create a differentiable model fiting the input geometries and configurations of our metasurface scatterers to the low-dimensional representation of the output field. 
To validate our model, we inverse design two optical elements: a wavelength multiplexed element that focuses light for $\lambda=633$~nm and produces an annular beam at $\lambda=400$~nm and an extended depth of focus lens.
\end{abstract}

\section{Introduction}
Controlling the optical response of a scatterer via its geometry is the fundamental goal of nanophotonics. 
In recent years, devices consisting of periodically arranged sub-wavelength structures, each of which can be engineered to scatter light, have shown promising results in both miniaturizing existing optics~\cite{Arbabi2015,Wang2018} as well as in creating elements with new electromagnetic (EM) properties~\cite{Zhan2017,Arbabi2017, Faraon2015,Colburneaar2114}. 
While such devices, also known as metasurfaces, do provide an extremely large number of parameters for designing optics, it is often challenging to harness all of these degrees of freedom relying solely on intuition. 
Adapted from the fluid dynamics community, inverse design provides an alternative paradigm to solve this problem. 
Here, the quality of a device's performance is characterized by a mathematical figure of merit (FOM). 
The design method entails running a forward simulation of Maxwell's equations for a specific configuration of the scatterers to calculate the FOM, and back-propagating through the physical equations to optimize the FOM and, subsequently the device's performance, by updating the geometry. 
Thus, instead of a trial-and-error approach, the large design space is efficiently searched using sophisticated numerical optimization methods. 
Inverse design has gained considerable interest from the nanophotonics community~\cite{JelenaInverse}, and it has already been used to design photonic elements~\cite{JelenaInverse,Piggott2017,Piggott2015}, plasmonic nanostructures~\cite{Hansen15}, and metasurfaces~\cite{Pestourie:18,Bayati_2020,Lin:19,Zhan:18,ellipsoids2020,Zhaneaax4769}. 
However, inverse design requires running the forward simulation many times, and thus the ultimate speed of the design depends directly on the computational efficiency of the forward simulation.

In most existing design methods, Maxwell's equations are solved on a meshed grid and refractive indices are allowed to change at every point on this grid~\cite{ Pestourie:18,Bayati_2020,Lin:19,Hansen15,JelenaInverse,Piggott2017,Piggott2015,Mansouree:20}. 
These methods are accurate and can simulate a scatterer of arbitrary shape; however, they do not scale well to large systems with small features, due to the time and memory required to carry out each forward simulation. 
In contrast, Mie-scattering-based analytical solutions~\cite{Zhan:18,ellipsoids2020,Zhaneaax4769} scale better in time and memory, as computational cost only depends on the number of scatterers in each device, but they only work for highly restrictive geometries. 
To date, only spherical~\cite{Zhan:18} and sparsely packed ellipsoids~\cite{ellipsoids2020} can be simulated using Mie scattering for inverse design. 
Another option is to rely on the local phase approximation (LPA)~\cite{lpa2017}, which assumes the EM properties of a single scatterer can be characterized as a single complex-valued transmission coefficient that is a function of the scatterer geometry, and independent from the geometries of the nearby scatterers. 
This approach requires solving Maxwell's equations under Bloch boundary conditions with methods such as rigorous coupled wave analysis (RCWA)~\cite{S4}. 
Such methods are fast but inaccurate when scatterers on a metasurface are not identical, which is especially apparent when geometries vary rapidly in space, or when scatterers are made from materials with low refractive indices. 

The objective of this work is to create a forward simulation method for inverse design that is faster than grid based methods, is not restricted to spheroidal particles, and is more accurate than methods relying on LPA.  
We will leverage several data-driven modeling and machine learning techniques~\cite{Brunton2019book}, which are being adopted in the field of optics and photonics~\cite{won2018intelligent}, with examples in fiber lasers~\cite{Fu2014oe,Brunton2014ieeejstqe,andral2015fiber,woodward2016towards,baumeister2018deep,sun2020deep} and metamaterial antennas~\cite{Johnson2015josaa}.

The EM response $\vec{\mathcal{E}}$ to a incident current $\vec{J}$ is given by Maxwell's equation:
\begin{equation}
\nabla \times \nabla \times \vec{\mathcal{E}}(\textbf{x}) - \omega^2 \epsilon(\textbf{x}) \mu(\textbf{x}) \vec{\mathcal{E}}(\textbf{x}) + i \omega \mu(\textbf{x}) \vec{J}(\textbf{x}) = 0
\end{equation}
where $\omega$ is the angular frequency of the current source, $\epsilon(\textbf{x})$ is the dielectric permittivity distribution, $\mu(\textbf{x})$ is the magnetic permeability distribution (assumed to be unity here as we will use dielectric materials) and the vector $\textbf{x}$ is the position vector. This implies that the field response $\vec{\mathcal{E}}(\textbf{x})$ only depends on the distribution of $\epsilon(\textbf{x})$. A forward simulation of Maxwell's equation thus entails the prediction of the spatial EM modes as a function of the scatterer geometry and position. Here, we first use high-fidelity EM simulations to generate data, which are then used to find a simple mapping between $\epsilon(\textbf{x})$ and $\vec{\mathcal{E}}(\textbf{x})$ exploiting the singular value decomposition (SVD) and neural networks~\cite{brunton_kutz_2019}. We note that, a number of previous works used neural networks to predict the spectral responses from metallic~\cite{Malkiel2018,Li:19,Ma2018} and dielectric~\cite{Peurifoyeaar4206,kiarashinejad2019,Deeplearningenabledinversedesigninnanophotonics,Liu2018,GaoLi2019,Lin:19,an2019novel} scatterers of various geometries. 
In these problems, the unit cells are identical, and hence there is no need to invoke LPA. 
However, for imaging applications, where the unit cells are spatially varying, invoking LPA results in inaccuracy. 
Our work aims to mitigate this challenge by using a data-driven framework to predict the spatial responses from dielectric circular cylinders, while including the effects of their nearest neighbors. Another recent work has applied data driven techniques to accelerate iterative finite difference frequency domain (FDFD) solvers~\cite{Trivedi2019}. 
While accurate, this method is however still memory intensive. 
Our work provides an alternative, interpolative method for simulating field responses from electromagnetic scatterers by fitting a differentiable model that maps the geometry of the scatterer and its closest neighbors to its EM field response. This model speeds up our forward simulation by estimating local patches of the EM field from the radius of a cylindrical scatterer and its surrounding neighbors. We found that this method can simulate a mesh with $1.2$ million discrete points $10^4$ times faster than conventional grid-based solvers, and is memory inexpensive enough that it can be run on a laptop.
We use this framework to inverse design two devices, both of which are unintuitive under the forward methodology: a multi-wavelength metasurface that produces a annulus beam for one wavelength, and focuses light at a different wavelength, as well as an extended depth of focus lens. 

\section{Methods}
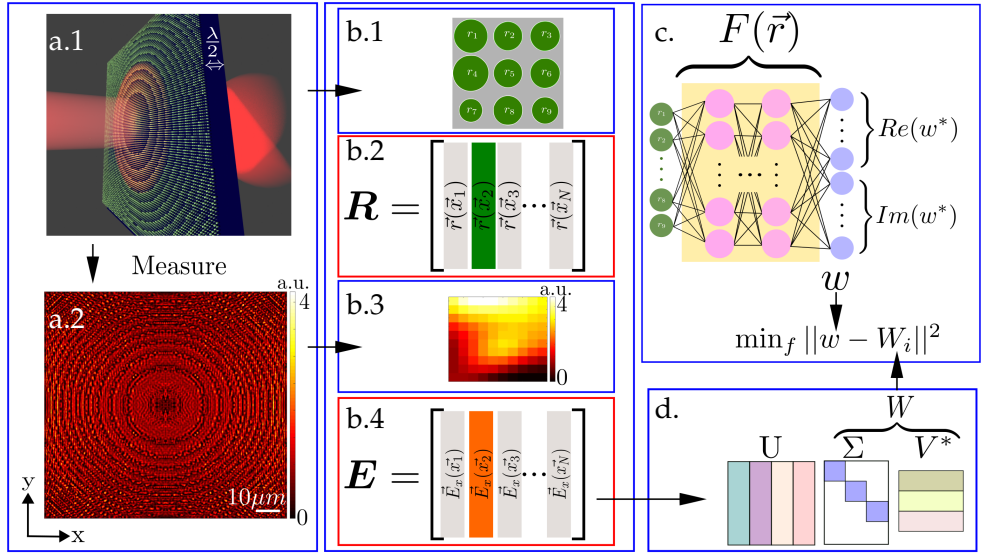
\begin{figure}[t]
\center
\begin{overpic}[width=.8\textwidth]{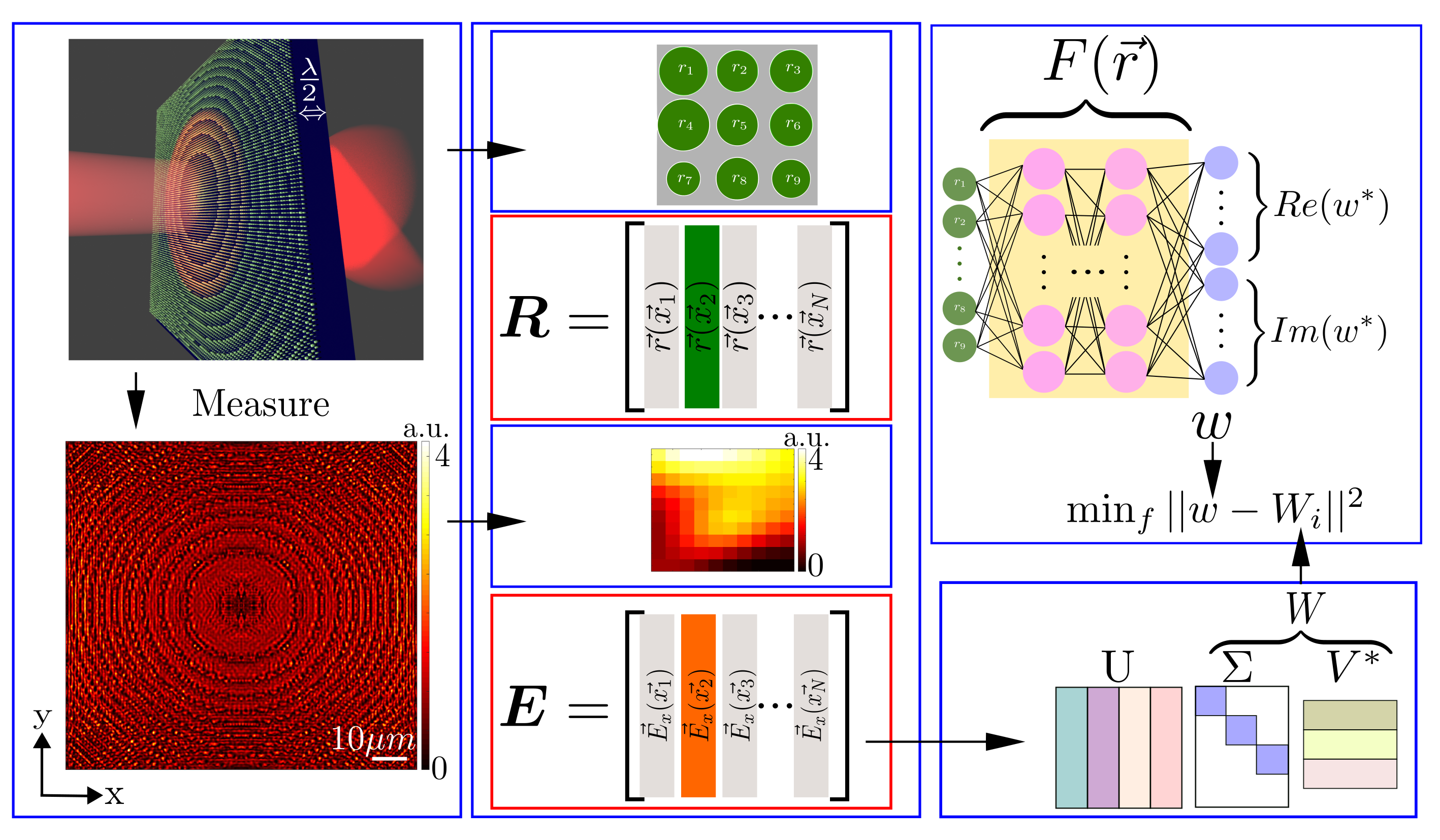}
 \put (5,52) {\textcolor{white}{a.1}}
 \put (5,24) {\textcolor{white}{a.2}}
 \put (35,53) {{b.1}}
 \put (35,41) {{b.2}}
 \put (35,25.5) {{b.3}}
 \put (35,14) {{b.4}}
 \put (66,53) {c.}
 \put (66,15) {d.}
\end{overpic}
\caption{Overview of method. \textbf{a.1.} Sample forward designed metasurface. \textbf{a.2.} Near-field response of metasurface for $\lambda =633$~nm \textbf{b.1-4} Parsing the data. \textbf{b.1.} Iterate through each pillar except the ones in the edges, and gather the surrounding pillar radii. \textbf{b.2.} Pillar radii and recorded and stacked into matrix $\bold{R}$. \textbf{b.3.} Field response in a square region with dimension of the pitch $p$ corresponding to the central pillar. \textbf{b.4} Electric fields are vectorized and stacked into a matrix $\bold{E}$. \textbf{c.} We create a neural network that predicts a vector $\bold{w}$ corresponding to the column of matrix \textbf{d.} $\bold{W}$ is constructed as the product $\bold{\Sigma} \bold{V^*}$, where $\bold{\Sigma}$ and $\bold{V}$ are taken from the SVD of $\bold{E}$.}
\label{Fig1}
\end{figure}

The goal of this work is to develop a fast and accurate proxy for the forward simulation that is differential and may be used for inverse design.  
Fig.~\ref{Fig1} shows the schematic of our strategy to build a differentiable map $F: \mathbb{R}^9 \rightarrow \mathbb{C}^{100}$ that predicts the electric field over a square area with dimension $p$ from the dielectric permittivity distribution $\epsilon(\textbf{x})$, modeled as 9 cylinders. Here $p$ is the periodicity of the metasurface, and each square area (unit cell) has been discretized into a $10\times10$ grid. The corresponding field being predicted is flattened into a $100 \times 1$ vector. We will explore two models: a low-dimensional linear regression model based on the singular value decomposition (SVD) and a deep neural network model to fit $F$. 

\subsection{Training data}
To train these models, we first generate a data set consisting of forward simulations of several physical devices, in our case lenses. The intuition is that the lens is arguably the simplest physical device, and will likely provide a useful basis to interpolate future devices.  
We forward designed $10$ lenses of diameter $\sim 50\mu m$ with focal lengths varying from $10-100\,\mu m$ ~\cite{msreview}. The lens design parameters are summarized in Table~\ref{table1}. All lenses are intended to function with a current source wavelength $\lambda = 0.633\,\mu m$. 
The material refractive index was set to $n=2$, corresponding to silicon nitride, our material of choice for visible wavelength operation~\cite{Colburn:18}.
These dimensions correspond to exactly 113 pillars on each axis of the metasurface. 
All the scatterers were computed with RCWA package S4~\cite{S4}. 
A sample lens of focal length $f=50\,\mu m$ is shown in Fig.\ref{Fig1}.a.1. 
We simulated the EM response of each lens using an x-polarized plane wave ($\lambda=0.633\,\mu m$) with the field monitor $\lambda/2$ away from the scatterers using Lumerical finite difference time domain (FDTD) software. An example field is shown in Fig.\ref{Fig1}.a.2. 
Only the $\mathcal{E}_x$ component was recorded due to minimal contributions to the total field power from other vector components, which is a result of the circular symmetry of the scatterers. 
However, the process could easily be generalized to predict the entire vector-field. 
The resolution of the simulation was chosen to be $0.04431 {\,\mu m}/{vox}$ in order to balance computational time, memory requirements, and accuracy. This results in (10$\times$10) field points in each square unit cell with dimension $p=443nm$ corresponding to each pillar.

\begin{table}
\caption{Parameters used to forward design the training data-set. $f$: focal length; $h$: height of the pillars; $n$: material index; $\lambda$: current source wavelength. Lens diameter $D$ is chosen to be the closest integer multiple of the periodicity $p$.}
\begin{center}
 \begin{tabular}{c c c c c c c} 
 \hline
 Parameter & $f$ & $D$ & $p$ & $h$ & $n$ & $\lambda$\\\hline
 Value & $10-100\,\mu m$ 
 & $50.0703\,\mu m$ 
   & $0.4431\,\mu m$
  & $0.633\,\mu m$ 
 & 2
  & $0.633\,\mu m$\\
 \hline
\end{tabular}
%  \begin{tabular}{|c| c|} 
%  \hline
%  Parameter & Value \\ [0.5ex] 
%  \hline
%  $f$ & $10-100\,\mu m$ \\ 
%  \hline
%  $D$ & $50.0703\,\mu m$ \\
%  \hline
%   $p$ & $0.4431\,\mu m$ \\
%  \hline
%  $h$ & $0.633\,\mu m$ \\
%  \hline
%  $n$ & 2\\
%  \hline
%  $\lambda$ & $0.633\,\mu m$\\
%  \hline
% \end{tabular}
\label{table1}
\end{center}
\end{table}

Once all of the field data were gathered, we constructed two matrices: $\textbf{R} \in \mathbb{R}^{9 \times N}$ for the radii, and $\bold{E} \in \mathbb{C}^{100 \times N}$ for the electric fields, with $N$ being the number of scatterers. 
The matrix $\textbf{R}$ was created by iterating over pillar location $\textbf{x}_\textbf{i}$, and storing the radii of the pillar and its 8 nearest neighbors as a column vector, shown in Figs. \ref{Fig1} b.1 and \ref{Fig1}.b.2. 
The pillars on the edges of the metasurface do not have neighbors and were neglected. 
Similarly, the matrix $\textbf{E}$ was created by iterating over each pillar location, extracting the field in the unit cell with centroid $\textbf{x}_\textbf{i}$, and storing it as a flattened $100 \times 1$ column vector, shown in Figs \ref{Fig1}.b.3 and \ref{Fig1}.b.4. 
This results in two matrices having $N = (113-2)^2 \times 10 = 123210$ columns.

\subsection{Linear regression model}
\begin{figure}[t]
\center
\begin{overpic}[width=.935\textwidth]{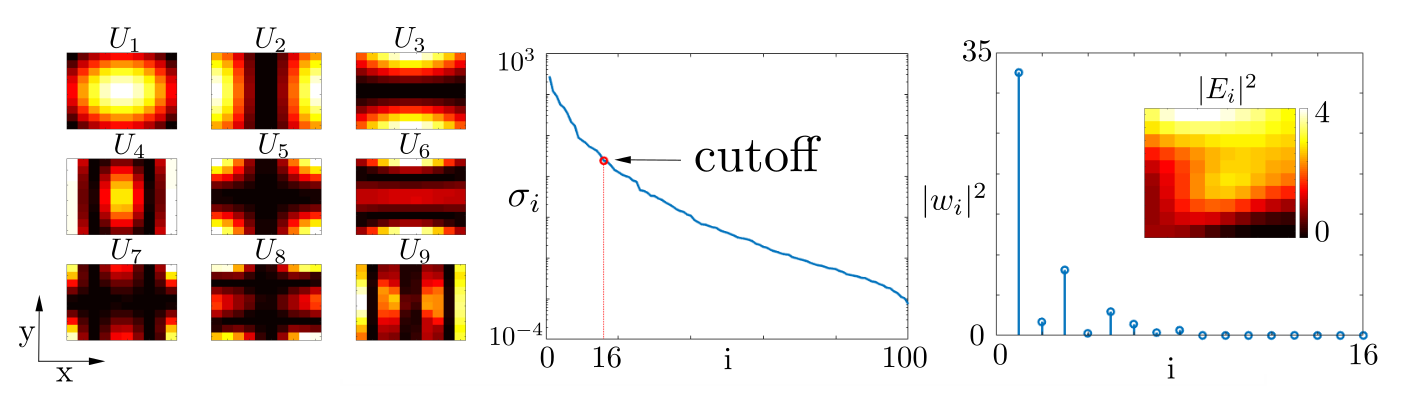}
\put(2,26.5) {a.}
\put (35,26.5) {b.}
\put (68.5,26.5) {c.}
\end{overpic}
\vspace{-.1in}
\caption{Singular value decomposition of simulated data. \textbf{a.} First 9 left hand singular vectors $\bold{U}$ of $\bold{E}$ matrix. \textbf{b.} Singular value decay of the diagonal matrix $\bold{\Sigma}$. Red circle represents the cut off order we used to reconstruct the electric fields. \textbf{c.} Plot of the absolute values squared of a random vector $\bold{(\Sigma V^*)_i} = \bold{w_i}$ that reconstructs some random $p\times p$ field. $\bold{w_i}$ represents the weights of the left hand singular vectors $\bold{U}$.}
\label{fig2}
\end{figure}

We first explore the low-dimensional structure of the matrix $\textbf{E}$, which will facilitate learning a map between the columns of $\textbf{R}$ and $\textbf{E}$.  
Patterns in the rows and columns of $\textbf{E}\in\mathbb{C}^{M\times N}$ are extracted via the singular value decomposition (SVD)~\cite{brunton_kutz_2019}: 
\begin{equation}
\textbf{E} = \textbf{U} \bold{\Sigma} \bold{V^*}
\end{equation}
where $\bold{U} \in \mathbb{C}^{M \times M}$ and $\bold{V} \in \mathbb{C}^{N \times N}$ are unitary matrices, and $\bold{\Sigma} \in \mathbb{R}^{M \times N}$ is a diagonal matrix, with non-negative entries on the diagonal and zeros off the diagonal. 
The columns of $\bold{U}$ can be thought of as a set of orthonormal basis vectors with which to represent the columns of $\bold{E}$. 
These columns of $\bold{U}$ are arranged hierarchically in terms of how much variance they capture in $\bold{E}$, as quantified by the corresponding diagonal element of $\bold{\Sigma}$.  
Fig.\ref{fig2} a. shows the square of the absolute value of the first 9 column vectors $\bold{u_j}$, reshaped from $\mathbb{C}^{100 \times 1}$ to $\mathbb{C}^{10 \times 10}$. 
Definite patterns are observed in these vectors, implying a low-dimensional representation of our data. 
The rows of $\bold{V^*}$ correspondingly provide a hierarchical basis for the rows of $\bold{E}$.  
Each column of the matrix $\bold{\Sigma V^*}$ determines the exact combination of the columns of $\bold{U}$ required to reproduce the corresponding column of $\bold{E}$. 
 
Guided by the SVD, it is possible to write an approximate matrix $\bold{\tilde{E}}$ as:
\begin{equation}\label{Eq:ESVDapprox}
\bold{\tilde{E} = U_q W}
\end{equation}
where $\bold{W} = \bold{\Sigma_q V^*_q}$, and the subscript $q < M$ is the truncation order of the matrix approximation.  The first $q$ columns of $\bold{U}$ are arranged to form $U_q$, the first $q\times q$ sub-block of $\bold{\Sigma}$ is extracted to form $\bold{\Sigma_q}$ and the first $q$ rows of $\bold{V^*}$ are taken to form $\bold{V^*_q}$. 
It can be shown that $\bold{\tilde{E}}$ is the best rank$-q$  approximation to the matrix $\bold{E}$, in the Frobenius norm~\cite{trefethen97}. 
We choose a truncation value of $q=16$, shown as the red circle in Fig.\ref{fig2}.b., as the rank $16$ approximation $\bold{\tilde{E}}$ captures $99\%$ of the variance in the matrix $\bold{E}$.

We will now construct a regression map to estimate columns of $\bold{E}$ from columns of $\bold{R}$.  Specifically, we estimate the matrix $\bold{W}$, which will be used to reconstruct $\bold{\tilde{E}}$.  
Instead of using the columns of $\bold{R}$ directly as features, we will create an augmented feature vector comprised of monomials constructed from the radii.  
This feature matrix $\bold{\Theta}$ is constructed by vertically concatenating integer Hadamard powers of $\bold{R}$:
% To predict the matrix $\bold{W}$, and reconstruct matrix $\tilde{E}$ with a linear model, we construct a feature matrix $\bold{\Theta}$ by vertically concatenating integer Hadamard powers of elements of $\bold{R}$:
\begin{equation}
\bold{\Theta} = \begin{bmatrix}
1\\
\bold{R}\\
\bold{R^{\circ 2}}\\
\vdots\\
\bold{R^{\circ m}}
\end{bmatrix}.% = \begin{bmatrix}
%1 & 1 & \dots & 1\\
%r_1(x_1) & r_1(x_2) & \dots & r_1(x_n)\\
%r_2(x_1) & r_2(x2) & \dots & r_2(x_n)\\
%\vdots & \vdots & \dots & \vdots \\
%r_9(x_1) & r_9(x_2) & \dots & r_9(x_n) \\
%r^2_1(x_1) & r^2_1(x_2) & \dots & r^2_1(x_n) \\
%r^2_2(x_1) & r^2_2(x_2) & \dots & r^2_2(x_n)\\
%\vdots & \vdots & \dots & \vdots \\
%r^2_9(x_1) & r^2_9(x_2) & \dots & r^2_9(x_n)\\
%\vdots & \vdots & \dots & \vdots \\
%r^p_1(x_1) & r^p_1(x_2) & \dots & r^p_1(x_n)\\
%r^p_2(x_1) & r^p_2(x_2) & \dots & r^p_2(x_n)\\
%\vdots &\vdots & \dots &\vdots\\
%r^p_9(x_1) & r^p_9(x_2) & \dots & r^p_9(x_n)
%\end{bmatrix}
\end{equation}
Where $\bold{R}^{\circ m}$ are the element-wise powers of $\bold{R}$:
\begin{equation}
R^{\circ m} = \begin{bmatrix}
r^m_{1,1}  & \dots & r^m_{1,N}\\
\vdots & \ddots & \vdots \\
r^m_{M,1} & \dots & r^m_{M,N}
\end{bmatrix}
\end{equation}
 Thus, we set up a linear system: 
\begin{equation}
\bold{W} = \bold{\Xi \Theta}
\end{equation}
and solve for $\bold{\Xi}$:
\begin{equation}
\bold{\Xi} \approx \bold{W \Theta^{\dagger}}
\end{equation}
where the superscript $^\dagger$ denotes the Moore-Penrose pseudo-inverse~\cite{trefethen97}. The matrix $\bold{\tilde{E}}$ can then be approximated by a \emph{generalized} linear regression problem:
\begin{equation}
\bold{\tilde{E}} \approx \bold{E_{\text{pred}} = \Xi \Theta}
\end{equation}

\begin{figure}[t]
\center
\begin{overpic}[width=\textwidth]{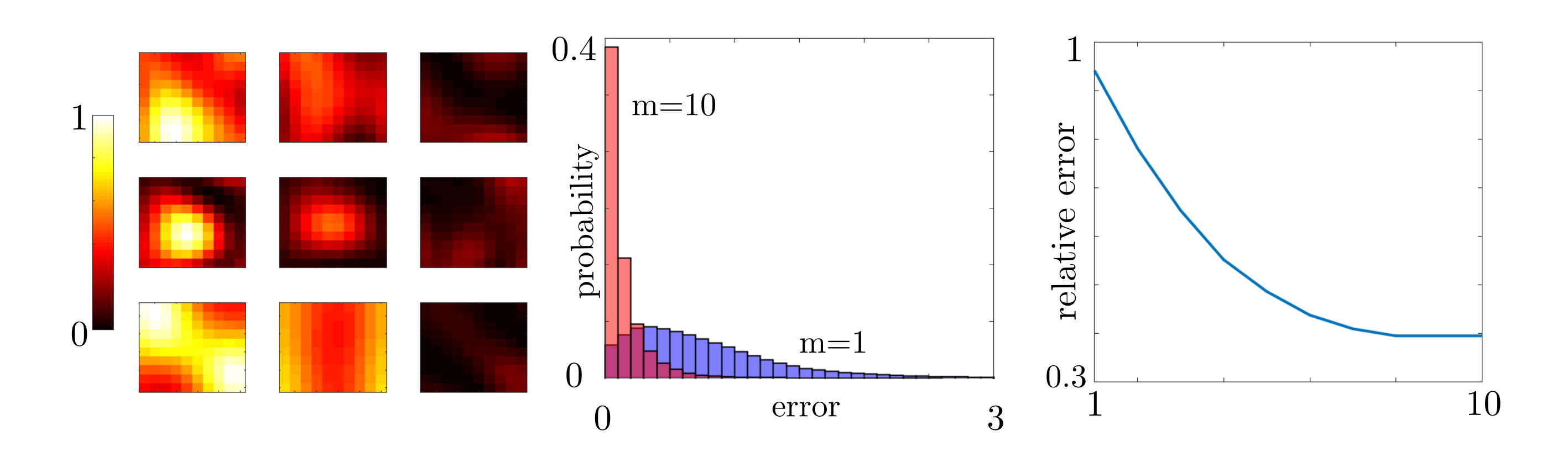}
\put(4.5,27) {a.}
\put (35,27.25) {b.}
\put (66.5,27.25) {c.}
 \put (11,27) {\scriptsize$|E_i|^2$}
 \put (18,27) {\scriptsize$|E_{\text{pred}}|^2$}
 \put (29,27) {\scriptsize$|\epsilon|^2$}
  \put (11,1) {I.}
 \put (20,1) {II.}
 \put (29,1) {III.}
 \put (81,2) {m}
\end{overpic}
\caption{\textbf{a.} Column \textbf{I.} represents the field simulated by FDTD, column \textbf{II.} is the electric field predicted using the linear model, and Column \textbf{III.} is the difference between the two. Each row represents the field corresponding to the same set of 9 radii. \textbf{b.} Probability density functions of relative errors between the predicted matrix $E_{\text{pred}}$ and the true matrix $\bold{E}$. The blue plot corresponds to a feature matrix with only $m=1$, and the red plot represents the feature matrix constructed with powers up to 10. \textbf{c.} Plot of the relative errors in the Frobenius norm between $\bold{E_{\text{pred}}}$ and $\bold{E}$. The x axis represents the power term in the radius features.} 
\label{Fig3}
\end{figure}
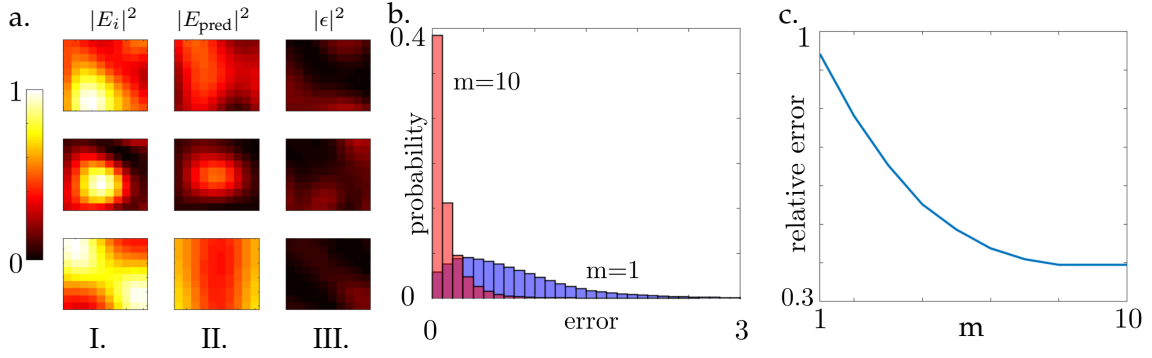

We varied the number of features used to train this linear model by changing the number of powers $m$ used to construct $\bold{\Theta}$. 
To train this linear model, we used $80\%$ of the data. 
After creating $\bold{W}$ and $\bold{\Theta}$, we extracted a random set of 98568 columns from each matrix in order to fit the matrix $\bold{\Xi}$. 
We used the other 24642 columns for validation. 
Fig.~\ref{Fig3}a depicts the qualitative performance of our linear model for $m=10$. 
Column \textbf{I} represents a randomly chosen vector $\bold{E_i}$ at some position $\bold{x_i}$, column \textbf{II} is the corresponding predicted vector $\bold{e_{\text{pred}}} \in \bold{E_{\text{pred}}}$, and column \textbf{III} is the absolute difference squared $|\epsilon|^2 = |\bold{E_i} - \bold{e_{\text{pred}}}|^2$. 
All values were normalized to have a maximum absolute value of $1$ and the same colorbar across all figures. 
Fig.~\ref{Fig3}b shows the probability density functions (PDFs) of the error distributions $|\bold{E_{\text{pred}}} - \bold{E}|^2$ as we increase $m$ from $1$ to $10$. 
As the number of features increases, the PDF becomes tighter. 
Fig. \ref{Fig3}c is a plot of the relative error defined as:
\begin{equation}\label{Eq:Error}
\frac{||\bold{E_{\text{pred}}}-\bold{E_{\text{\text{test}}}}||_F^2}{||\bold{E_{\text{\text{test}}}}||^2_F}
\end{equation}
as a function of $m$. Both plots show the error between our model and the FDTD simulation decreasing as the number of features in each matrix increases, so the model converges to the actual physics of the system. The final relative error for the linear model converged to $\approx 0.395$. As a side note, we have attempted using monomial expansions up to order $2$ of column vectors of $\bold{R}$ as input features for our model, by using powers of column vectors of $\bold{R}$ and cross terms in between radii, but found no significant improvements in relative error when fitting the model.

\subsection{Neural network model}
\begin{figure}[t]
\center
\begin{overpic}[width=\textwidth]{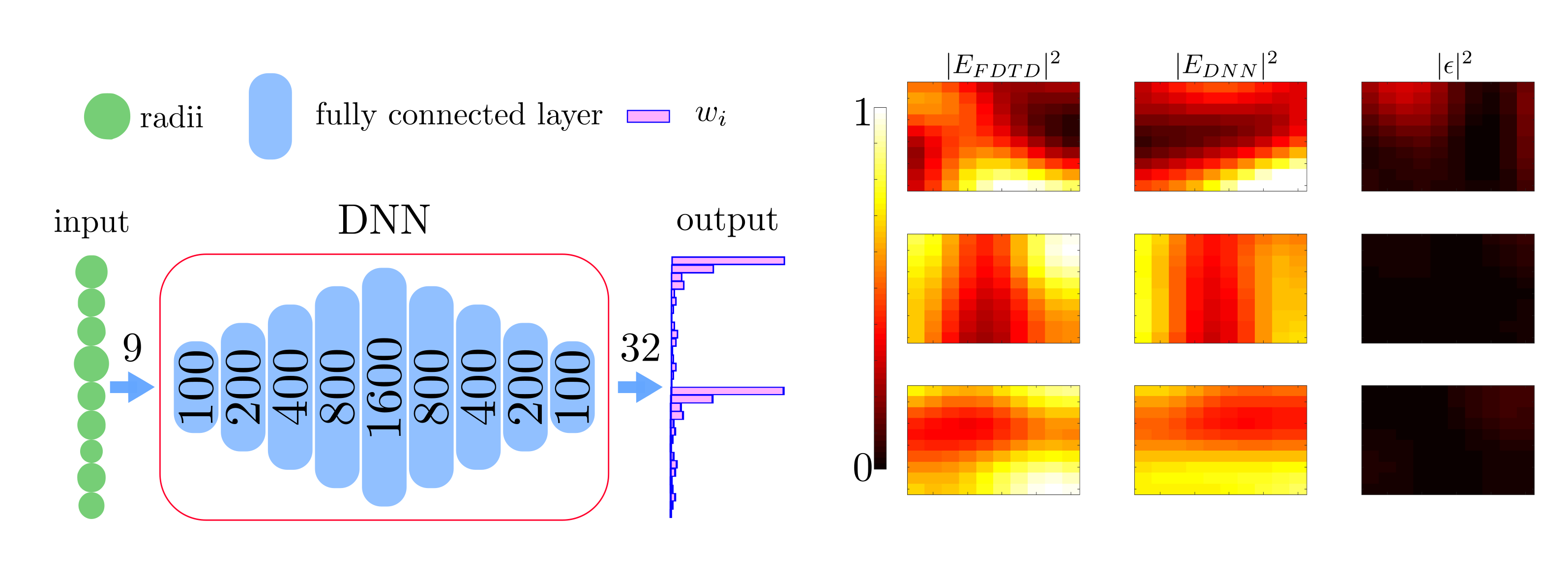}
 \put (2,33) {a.}
 \put (53,33) {b.}
 \put (63,3) {I.}
 \put (77,3) {II.}
 \put (91,3) {III.}
 \end{overpic}
\caption{\textbf{a.} The input into the DNN is 9 radii. The DNN architecture consists of 9 fully connected layers. The first layer starts off with 100 neurons, and each subsequent layer doubles the number of neurons until 1600, then number of neurons per layer is halved until the final layer has 100 neurons. All layers are followed by a ReLU activation function. The output has 32 elements. \textbf{b.} The performance of the DNN model. Column \textbf{I} is the field simulated by FDTD. \textbf{II.} is the field reconstructed by Eq.~\eqref{Eq:ESVDapprox} from the predicted vector $\bold{w_i}$. \textbf{III.} Difference between fields.}
\label{fig4}
\end{figure}
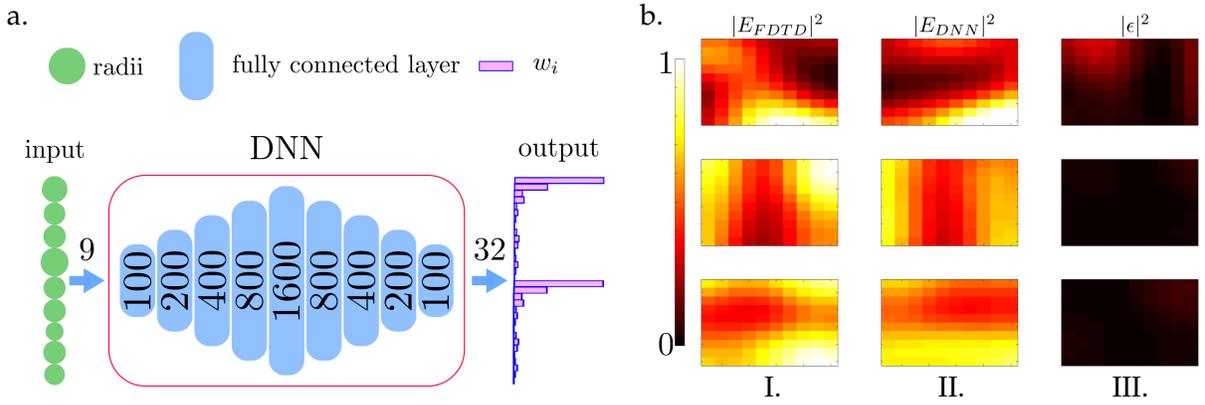

To improve on the generalized linear model, we construct a deep neural network (DNN), shown in Fig.~\ref{fig4}. 
We hypothesize that the DNN would learn a non-linear transformation of the input features that better capture the physics of the system. 
The model was trained by using $80\%$ of the data set, while keeping $20\%$ for validation. 
The architecture was implemented in TensorFlow~\cite{tensorflow2015-whitepaper} and optimized using the Adam optimizer~\cite{kingma2014adam}. 
The DNN architecture consists of 11 fully connected layers, each followed by a ReLU activation function. 
The first layer of the network is the input layer with 9 neurons corresponding to each radius. 
The 2nd layer has 100 neurons, which was doubled with each subsequent layer until 1600, and then cut in half until the second to last layer again had 100 neurons. The final layer had 32 neurons, with the first 16 elements corresponding to the real components of the vector $\bold{w}$ and the last 16 components corresponding to the imaginary components of $\bold{w_i}$. The outputs were arranged in this manner due to TensorFlow's limitations when designing complex-valued neural networks. The objective function used was a mean squared error between the output vector, and the corresponding vector from $\bold{W}$, shown in detail in Figs. \ref{Fig1}c. and \ref{Fig1}d. The network was trained until the mean squared error of the verification data set stopped being minimized in order to avoid over-fitting. Once the network was trained, we computed the electric field response of the training data by feeding the test data set into the neural network to compute $\bold{W_{\text{pred}}}$, and Eq.~\eqref{Eq:ESVDapprox} to compute $\bold{E_{\text{pred}}}$. 
The quality of our prediction can again be summarized by the relative error between $\bold{E_{\text{pred}}}$ and $\bold{E_{\text{test}}}$ in the Frobenius norm, given in Eq.~\eqref{Eq:Error}, which was computed to be
$\approx 0.26$. 
We use the DNN model to inverse design our devices.

\section{Results}
To test the utility of our model, we inverse designed two meta-optical devices. Motivated by stimulated emission depletion (STED) microscopy~\cite{Vicidomini2018}, the first device we inverse designed was a wavelength multiplexed lens that focuses light with $\lambda=633$~nm, and creates a annulus beam at the focal plan for $\lambda=400$~nm.  The second is an extended depth of focus (EDOF) lens that focuses light over $100-350\,\mu m$ along the optical axis. The optimization process was implemented in TensorFlow~\cite{tensorflow2015-whitepaper}. We used the DNN model and Eq.~\eqref{Eq:ESVDapprox} to predict the nearfields of the designed devices. The farfields were then calculated by using the angular propagation method~\cite{goodman}. The gradients with respect to radii were calculated by using TensorFlow's auto-differentiation, and updated by the adam optimizer.

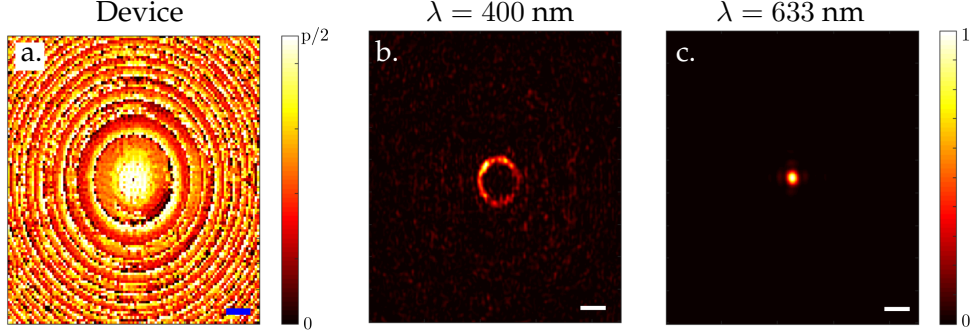
\begin{figure}[t]
\center
\begin{overpic}[scale=0.95]{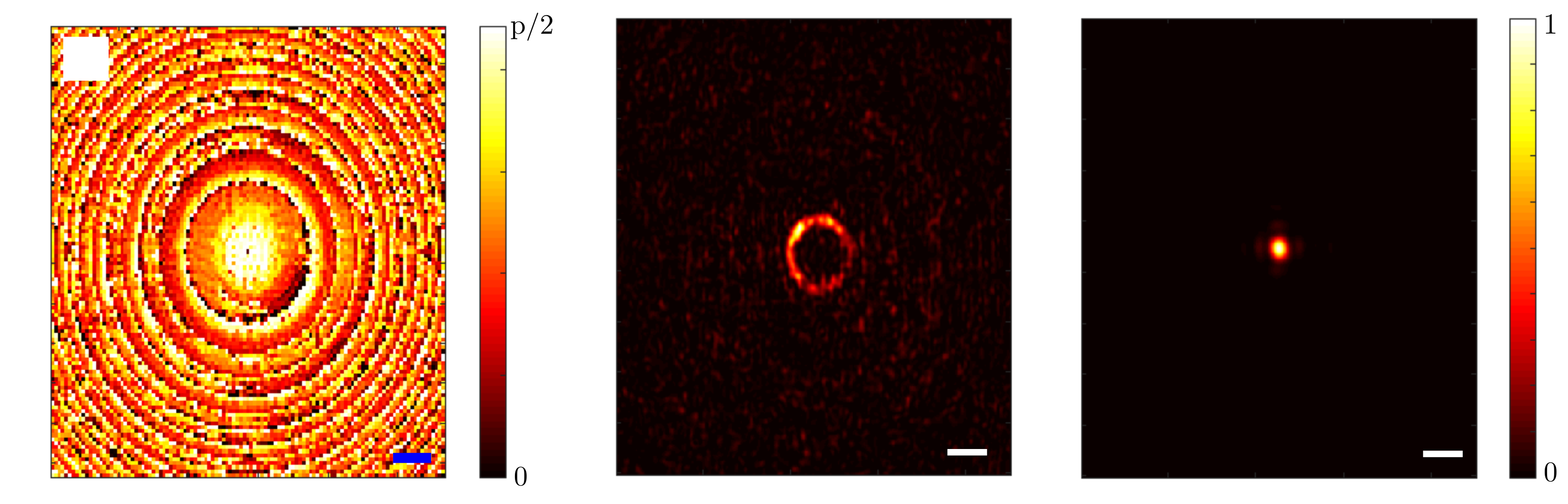}
\put (12,32) {Device}
 \put (45,32) {$\lambda = 400$~nm}
 \put (74,32) {$\lambda = 633$~nm}
 \put (4.5,28) {a.}
 \put (40,28) {\textcolor{white}{b.}}
 \put (70,28) {\textcolor{white}{c.}}
\end{overpic}

\caption{\textbf{a.} Optimized multi-functional device. Scale bar is $5 \mu m$ \textbf{b.} FDTD result for $\lambda = 0.4 \,\mu m$. \textbf{c.} FDTD result for $\lambda = 0.633 \,\mu m$. Scale bars are $2 \,\mu m$. Units are normalized so the maximum intensity is equal to 1.}
\label{fig5}
\end{figure}

To design the multi-wavelength lens, we had to predict nearfields for two wavelength. Hence, we repeated the procedure outlined in section 2 to create one more data driven model to predict the field response for a $\lambda = 400$~nm current source. Once trained, the relative error defined by Eq.~\eqref{Eq:Error} computed on the test dataset for $\lambda = 400$ nm was $\approx 0.37$. The difference between the two wavelengths can be explained by the relatively non-smooth transmission of $\lambda=400$ nm E-fields over this range of radii when compared to the $\lambda = 633$ nm case. To optimize the lens, we defined two figures of merit for each wavelength as:
\begin{align}
\text{FOM}_{400} &= - \sum_{m= 0}^{19} I\left(c \cos\left(m \frac{2 \pi}{20}\right), c \sin
\left(m \frac{2 \pi}{20}\right), 50 \,\mu m\right) \\
\text{FOM}_{633} &= -20 \times I(0,0, 50 \,\mu m) ,
\end{align}
Where the function $I(x,y,z)$ is the intensity of the electric field at $(x,y,z)$ coordinate given by $I(x,y,z) = \mathcal{E}^*(x,y,z) \mathcal{E}(x,y,z)$. The constant $c = 1.5 \,\mu m$ corresponding to the radius of the annular beam at the focal spot. The tuple $(c \cos(i \frac{2 \pi}{20}), c \sin(i \frac{2 \pi}{20}))$ is the parametrization of a circle in the $x-y$ plane, that we discretized over 20 points on the circle. The factor of $20$ on $FOM_{633}$ is chosen as a normalization factor to ensure the integral of the intensity over the annulus is the same as the intensity at focal spot. The quantity optimized was then:
\begin{equation}
\max(\text{FOM}_{400},\text{FOM}_{633})
\end{equation}
with respect to the radii distribution. We set our initial radius distribution to be the same as the forward designed lens for $\lambda = 633$~nm and $f=50\,\mu m$. The designed device is shown in Fig. \ref{fig5}a. To verify the design, we computed the nearfield response of the radii distribution in Lumerical FDTD and propagated the nearfield to the focal plane using angular spectrum method. Figs. \ref{fig5}b and \ref{fig5}c show the meta-optic's response to $400$~nm and $633$~nm wavelength at the focal plane respectively. The efficiency $\eta$ of the metasurface was calculated to be $26.82\%$ for $\lambda = 400 nm$. The formal definition is given in the appendix. We quantify the annulusal functionality of the metasurface as the ratio between the power confined in the annulus to the power confined in the center of the annulus. This ratio $\eta_\circ$ was calculated to be $58.47$ for $\lambda = 633nm$, formally defined in the appendix.
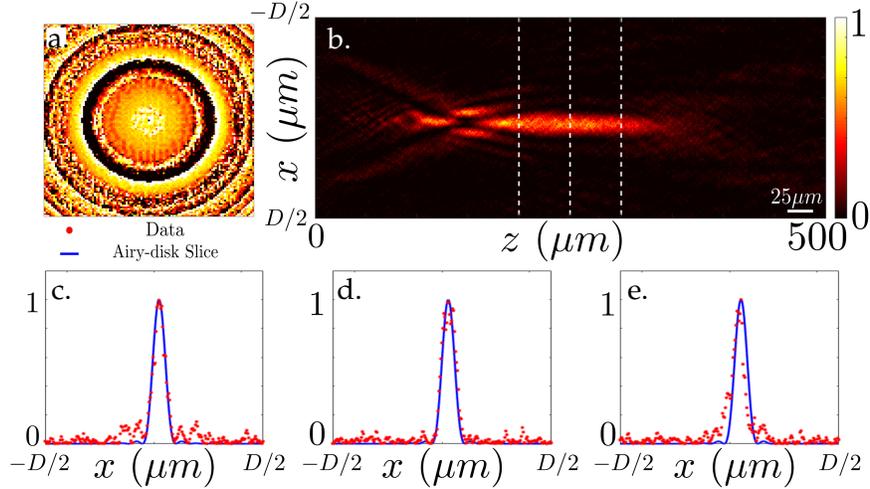
\begin{figure}[t]
\center
\begin{overpic}[scale=0.95]{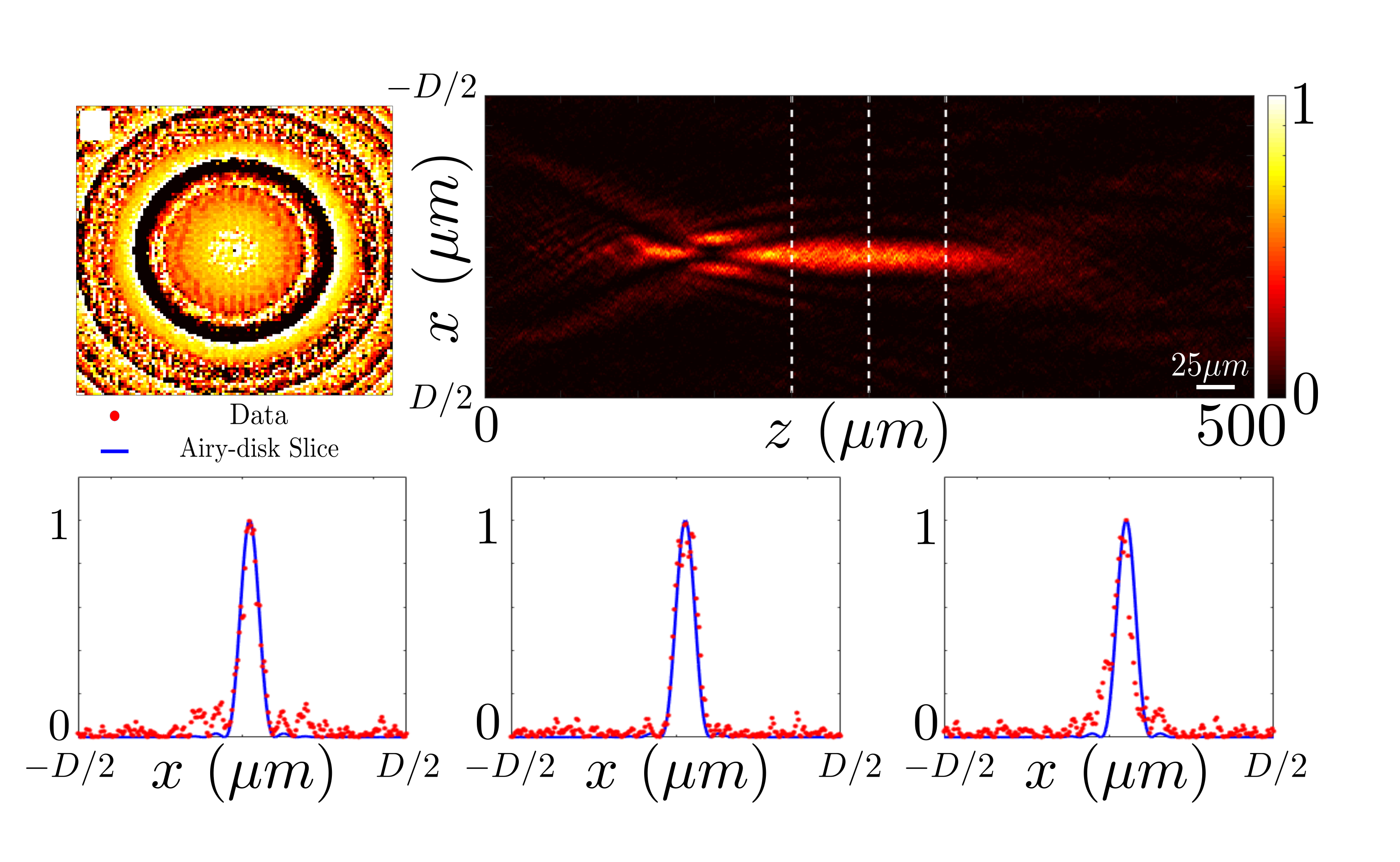}
 \put (5.85,51.6) {\textcolor{black}{a.}}
 \put (36,51.5) {\textcolor{white}{b.}}
 \put (6.25,24.5) {\textcolor{black}{c.}}
 \put (37,24.5) {\textcolor{black}{d.}}
 \put (68.25,24.5) {\textcolor{black}{e.}}
\end{overpic}

 \vspace{-.2in}
\caption{EDOF lens. \textbf{a.} Device. \textbf{b.} Simulated field. Simulation is a result of FDTD and angular spectrum propagation. \textbf{c-e} Slices of field in \textbf{b.} along the dashed lines corresponding to $200\,\mu m$, $250\,\mu m$ and $300 \,\mu m$ respectively. The red dots are the simulated data. The blue lines are the Airy disk profiles corresponding to the diffraction limit. All intensities are normalized by the corresponding maximum intensity.}
\label{edof}
\end{figure}

The EDOF lens was designed by using a lens with $f=100\,\mu m$ as a starting condition. Our intent was to design an EDOF lens to focus from $50\,\mu m$ to $100\,\mu m$. We defined the figure of merit for the EDOF as:
\begin{equation}
\text{FOM}_\text{EDOF} = -\sum_{m=0}^{10} \log(I(0,0,50+m\times dz))
\end{equation}
where $dz = 10\,\mu m$. Thus we aim to maximize the intensity at the center of $10$ equi-spaced $x-y$ planes. The resulting device is shown in Fig. \ref{edof}a. The intent behind the logarithmic sum was to equalize the importance of each term along the z-axis. Without the logarithm term, the optimization would prioritize a single focal spot, since such a device would minimize the figure of merit. After optimization, this figure of merit converged to a lens focusing from $100\,\mu m$ to $350 \,\mu m$ as shown in Fig. \ref{edof}b, corresponding to a numerical aperture varying from 0.07-0.25. We attribute the longer depth of focus than what was intended to the physical nature of wave propagation. Figs. \ref{edof}c-\ref{edof}e show slices of the electric field at $200\,\mu m$, $250\,\mu m$ and $300 \,\mu m$ along the optical axis. The red dots correspond the the simulated data, and the blue line corresponds to the Airy disk corresponding to the diffraction limited focal spot.  We note that clearly, there are additional side-lobes in the EDOF design, and thus the total energy in the main lobe suffers. However, a different figure of merit can be designed to reduce the side-lobes, depending on the desired application.
\section{Discussion}
This paper outlines a data driven methodology for forward simulation of Maxwell's equations to design optical metasurfaces. Our model does not make the local phase approximation, and thus the inter-scatterer coupling is well accounted for. While, the model is not as accurate as a complete full-wave simulation, it is significantly faster. A single forward simulation of a square area of dimensions $50\mu m \times 50 \mu m$ at $ 44.31$~nm resolution takes approximately 12 seconds with our method versus approximately 3.1 hours using Lumerical FDTD software in the same computer. FDTD also requires a 58.95 GB initialization mesh and 29.6GB of RAM for the same simulation, while our method only requires 3.75GB for the same problem, and can be run on a mid-range laptop.

It is worth noting that our method is inherently interpolative, and thus is only as accurate as the data that we feed into it. Therefore the current model is limited to predicting fields from lens-like devices. One way we could improve this model is by using additional data to train it. In our future work, we hope to improve the accuracy of this model by simulating random arrangements of scatterers and using this as our training data set in addition to the data set from lenses. We also emphasize that the reported efficiency of the designed lenses is low, which remains a challenge for low index materials \cite{Bayati19}. However, full wave simulations have reported efficiency increase of the metasurface lenses especially when all the coupling between scatterers are exactly accounted for \cite{Mansouree:20, Chung:20}. Our model could be improved by better accounting for the coupling between scatterers using more data, especially EM field responses from scatterers with rapidly varying geometries, since the scatterer geometries of lenses vary slowly in space. Modelling the second nearest neighboring scatterers could also be an interesting path forward. Furthermore, adding additional constraints such as assumptions about energy conservation to the model training process, could further increase the accuracy of the model.

\section{Acknowledgements}
This research was supported by NSF-1825308, the UW Reality Lab, Facebook, Google, Futurewei, and Amazon.
M. Z. is supported by an NSF graduate research fellowship. A. M. is partially supported by Washington Research Foundation distinguished investigator award. S. L. B. acknowledges support from the National Science Foundation (NSF HDR award \#1934292). 

%%%%%%%%%%%
%% BIBLIOGRAPHY
%%%%%%%%%%%
%\bibliographystyle{plain}
% \begin{spacing}{.9}
% \small{
% \setlength{\bibsep}{6.5pt}
% \bibliographystyle{unsrt}
% \bibliography{mybib}
%5 }
% \end{spacing}

 \section{Appendix}
 The standard definition of efficiency for a metasurface lens with a given focal length $f$ is:
\begin{align}\label{Eq:metalensEff}
\eta &= \frac{\int \int _\Omega \mathcal{E}^*(x,y,z=f)\mathcal{E}^*(x,y,z=f) dx dy}{\int \int_{x,y}\mathcal{E}^*(x,y,z=0)\mathcal{E}^*(x,y,z=0) dx dy}\\
\Omega &:= x^2+y^2 < (3 \times FWHM)^2
\end{align}
where $\Omega$ is the surface around the focal spot which we integrate over, and FWHM is the full width half maximum of a Gaussian fitted to the focal spot. 

We quantify the functionality of the annular metasurface as the ratio between the power confined in the annulus to the power confined in the center of the annulus. More formally we define $\eta_\circ$ as:
\begin{align}\label{Eq:annuluseff}
\eta_\circ &= \frac{\int \int_{\Omega_\circ} \mathcal{E}^*(x,y,z=F)\mathcal{E}^*(x,y,z=F) dx dy}{\int \int_{\Omega_.} \mathcal{E}^*(x,y,z=F)\mathcal{E}^*(x,y,z=F) dx dy}\\
\Omega_{\circ} &:= (r_{\circ}+\delta r)^2 < x^2+y^2 < (r_{\circ}+\delta r)^2\\
\Omega_{t} &:= x^2 + y^2 < \delta r ^2
\end{align}
Here, $\Omega_{\circ}$ is the surface representing the annulus and $\Omega_t$ is the surface representing the center of the annulus. $r_\circ$ is the radius of the annulus, defined in the optimization procedure as $1.5\,\mu m$. $\delta r$ is the thickness over which we integrate, which we define as $\delta r = \frac{1}{2}$FWHM calculated for the $\lambda = 633$~nm case. $\eta_\circ$ was found to be equal to $58.47$. All integrals are taken over the $\lambda = 400$~nm field. Another possible metric of interest would be the fraction of power contained inside the surface $\Omega_t$, but in the $\lambda = 633$~nm case. We calculate this metric by switching $\Omega$ for $\Omega_t$ in Eq. \ref{Eq:metalensEff}. This metric gives an efficiency of $10.55\%$.

\end{document}